\documentclass[journal, final, 10pt]{IEEEtran}

\usepackage[T1]{fontenc}
\usepackage[utf8]{inputenc}
\usepackage{cite}
\usepackage{amsmath, mathtools}
\usepackage{amsfonts,amsthm,bm}
\usepackage{amssymb}
\usepackage{comment}
\usepackage{graphicx}
\usepackage{standalone}
\usepackage{dsfont}
\usepackage{mathtools}
\usepackage{color, colortbl}
\usepackage{soul}
\usepackage[normalem]{ulem}
\usepackage[acronym,shortcuts]{glossaries}
\usepackage{arydshln}
\usepackage{bbm}
\usepackage{svg} 
\usepackage{algorithm,algorithmic}
\usepackage{adjustbox}
\usepackage[inline]{enumitem}
\usepackage{multirow}
\usepackage{tabularx}
\usepackage{booktabs}

\usepackage{comment}
\usepackage{tikz}
\usepackage{pgfplots}
\pgfplotsset{compat=newest}
\pgfplotsset{plot coordinates/math parser=false}
\newlength\fheight
\newlength\fwidth
\usetikzlibrary{plotmarks,patterns,decorations.pathreplacing,backgrounds,calc,arrows,arrows.meta,spy,matrix}
\usepgfplotslibrary{patchplots,groupplots}
\usepackage{tikzscale}
\usepackage{balance}

\usetikzlibrary {patterns.meta}

\DeclareMathOperator*{\argmax}{\arg\!\max}

\DeclareMathOperator*{\diag}{diag}

\newcommand\blfootnote[1]{%
  \begingroup
  \renewcommand\thefootnote{}\footnote{#1}%
  \addtocounter{footnote}{-1}%
  \endgroup
}

\newacronym{3gpp}{3GPP}{3rd Generation Partnership Project}
\newacronym{5g}{5G}{fifth-generation}
\newacronym{6g}{6G}{sixth-generation}
\newacronym{af}{AF}{amplify-and-forward}
\newacronym{b5g}{B5G}{beyond fifth-generation}
\newacronym{cwc}{CWC}{capacity-weighted clustering}
\newacronym{da}{DA}{deterministic allocation}
\newacronym{dlos}{dLoS}{deterministic \ac{los}}
\newacronym{ecdf}{ECDF}{empirical cumulative distribution function}
\newacronym{fdma}{FDMA}{frequency division multiple access}
\newacronym{fpga}{FPGA}{field programmable gate array}
\newacronym{fov}{FoV}{field-of-view}
\newacronym{ga}{GA}{genetic algorithm}
\newacronym{gnb}{gNB}{next generation Node Base}
\newacronym{gmax}{GMAX}{Greedy MAXimum-rate scheduler}
\newacronym{hc}{HC}{hierarchical clustering}
\newacronym{hpbw}{HPBW}{half-power beamwidth}
\newacronym{km}{KM}{K-means}
\newacronym{kmed}{KMed}{K-medoids}
\newacronym{iab}{IAB}{integrated access and backhaul}
\newacronym{icwc}{ICWC}{inverse capacity-weighted clustering}
\newacronym{irs}{IRS}{intelligent reflecting surface}
\newacronym{isd}{ISD}{inter-site distance}
\newacronym{los}{LoS}{line-of-sight}
\newacronym{lsfc}{LSFC}{large-scale fading coefficient}
\newacronym{minlp}{MINLP}{mixed integer nonlinear programming}
\newacronym{mimo}{MIMO}{multiple input multiple output}
\newacronym{mmwave}{mmWave}{millimeter wave}
\newacronym{nlos}{NLoS}{non-line-of-sight}
\newacronym{noma}{NOMA}{non-orthogonal multiple access}
\newacronym{ofdma}{OFDMA}{orthogonal frequency-division multiple access}
\newacronym{oscbc}{OSCBC}{one-shot capacity-based clustering}
\newacronym{oma}{OMA}{orthogonal multiple access}
\newacronym{pam}{PAM}{partition around medoids}
\newacronym{plos}{pLoS}{probabilistic \ac{los}}
\newacronym{rb}{RB}{resource block}
\newacronym{rsma}{RSMA}{rate-splitting multiple access}
\newacronym{scm}{SCM}{spatial channel model}
\newacronym{snr}{SNR}{signal-to-noise-ratio}
\newacronym{sinr}{SINR}{signal-to-interference-plus-noise-ratio}
\newacronym{siso}{SISO}{single input single output}
\newacronym{svd}{SVD}{singular value decomposition}
\newacronym{tdma}{TDMA}{time division multiple access}
\newacronym{tti}{TTI}{transmission time interval}
\newacronym{thz}{THz}{Terahertz}
\newacronym{uoscbc}{UOSCBC}{unconstrained capacity-based clustering}
\newacronym{ue}{UE}{user equipment}
\newacronym{ula}{ULA}{uniform linear array}
\newacronym{uma}{UMa}{urban macro-cell}
\newacronym{umi}{UMi}{urban micro-cell}
\newacronym{upa}{UPA}{uniform planar array}

\title{Scheduling for Downlink OFDMA With IRS Reconfiguration Constraints}
\author{Alberto Rech,  Leonardo Badia,  Stefano Tomasin\\Department of Information Engineering, University of Padova, Italy.}
\begin{document}
\maketitle



\begin{abstract}
The technical limitations of the \ac{irs} (re)configurations in terms of both communication overhead and energy efficiency must be considered when \acp{irs} are used in cellular networks.
In this paper, we investigate the downlink time-frequency scheduling of an \ac{irs}-assisted multi-user system in the \ac{ofdma} framework wherein both the set of possible \ac{irs} configurations and the number of \ac{irs} reconfigurations within a time frame are limited.
We formulate the sum rate maximization problem as a non-polynomial (NP)-complete generalized multi-knapsack problem. A heuristic greedy algorithm for the joint \ac{irs} configuration and time-frequency scheduling is also proposed. 
Numerical simulations prove the effectiveness of our greedy solution. 
\end{abstract}

\begin{picture}(0,0)(0,-300)
\put(0,0){
\put(0,0){\qquad \qquad \quad This paper has been submitted to IEEE for publication. Copyright may change without notice.}}
\end{picture}


\glsresetall
\IEEEpeerreviewmaketitle
    
\section{Introduction}
\label{sec:introduction}

\Acp{irs} consist of meta-surfaces with radiating elements that can passively tune the phase shift of incoming signals to collectively reflect them in the desired propagation direction without active amplification~\cite{liu2021reconfigurable}.  \acp{irs} are considered among the most promising solutions to enhance the network coverage in challenging propagation conditions, e.g., for communications in the\ac{mmwave} band.

Downlink scheduling solutions for \ac{irs}-assisted communications have been extensively studied for cellular networks under several implementation constraints. 
Dynamic optimization schemes adjusting \ac{irs} configurations over each time slot have been explored in~\cite{Yang20IRS, Lee23Harmony}. 
Instead, the authors of~\cite{guo2021intelligent} consider a 2-user downlink transmission in a \ac{irs}-aided scenario over fading channels, comparing results of different basic \ac{oma} and \ac{noma} schemes. The study reveals that while \ac{noma} appears to be the best solution, \ac{tdma} outperforms \acl{fdma} due to the lack of frequency selective beamforming capabilities at the \ac{irs}. 
Additionally, the performance of \ac{noma} scheduling solutions, including \acl{rsma}, has been evaluated in~\cite{Bansal21Rate, Zhuo22Partial}.

Nevertheless, the majority of the literature on \acp{irs} relies on problematic assumptions. Specifically, the assumption of an ideal control channel with the base station is prevalent in the literature, while actual deployments are expected to have wireless, error-prone \ac{irs} control channels, possibly implemented with low-cost technologies~\cite{liaskos2018realizing}. 
This introduces constraints on the \ac{irs} reconfiguration period, which results in synchronization issues and increased power consumption~\cite{flamini2022towards}.
Indeed, early \ac{irs} prototypes display non-negligible phase-shift reconfiguration times \cite{rossanese2022designing, yezhen2020novel}. Such overhead increases with the size of the \ac{irs}, and it is expected to become a serious issue with the extremely large \acp{irs} needed to overcome channel losses in harsh propagation environments~\cite{nadeem2020asymptotic,jamali2022low}.
In this context, it is crucial to design resource allocation algorithms that mitigate the limitations imposed by the constrained \ac{irs} reconfigurations.
This kind of constraint has been studied in \cite{mu2021capacity}, with a characterization of both \ac{oma} and \ac{noma} schemes in a 2-user \ac{irs}-aided \acl{siso} system with Rayleigh fading channels. 
Furthermore, results for a \ac{tdma} scheduler in the multi-user \ac{mimo} case have been presented in \cite{rech2023downlinkJ}, where we propose several clustering techniques to optimize either the system sum rate or the user fairness.

In this paper, we propose an \ac{ofdma} scheduling policy for downlink cellular transmissions. We aim at maximizing the system sum rate by jointly performing resource allocation and \ac{irs} configuration.
The overhead from the \ac{irs} is constrained by limiting the number of reconfigurations that can be performed within each scheduling period, which forces the reuse of the same configurations for multiple users \cite{guidolin2014distributed}. Moreover, we also consider the case where the \ac{irs} configuration can only be chosen within a \textit{codebook} of configurations to provide a further reduction of the control overhead \cite{ghanem2023optimization}. 
We formulate the sum rate maximization problem as a non-polynomial (NP)-complete generalized multi-knapsack problem, and we propose a heuristic greedy algorithm for the joint \ac{irs} configuration and time-frequency scheduling. 
Numerical simulations prove the effectiveness of our greedy solution. 

The rest of the paper is organized as follows. In Section~\ref{sec:system_model} we introduce the \ac{ofdma} scheduling framework. In Section~\ref{sec:scheduling} we present the sum rate optimization problem and our proposed greedy scheduler. In Section~\ref{sec:numerical_results} we discuss the numerical results of our and the state-of-the-art schedulers. Finally, Section~\ref{sec:conclusions} concludes the paper.

\blfootnote{\noindent\emph{Notation.} Scalars are denoted by italic letters; vectors and matrices by boldface lowercase and uppercase letters, respectively; sets are denoted by calligraphic uppercase letters. $\bm{A}^{\rm T}$ and $\bm{A}^\dagger$ denote the transpose and the conjugate transpose of matrix $\bm{A}$, respectively. $\diag(\bm{a})$ indicates a square diagonal matrix with the elements of $\bm{a}$ on the principal diagonal. The imaginary unit is $j =\sqrt{-1}$. Finally, $\mathbb{E}[\cdot]$ denotes statistical expectation.}

\section{System Model}\label{sec:system_model}

\begin{figure}
    \centering
    \includegraphics[width=0.9\linewidth]{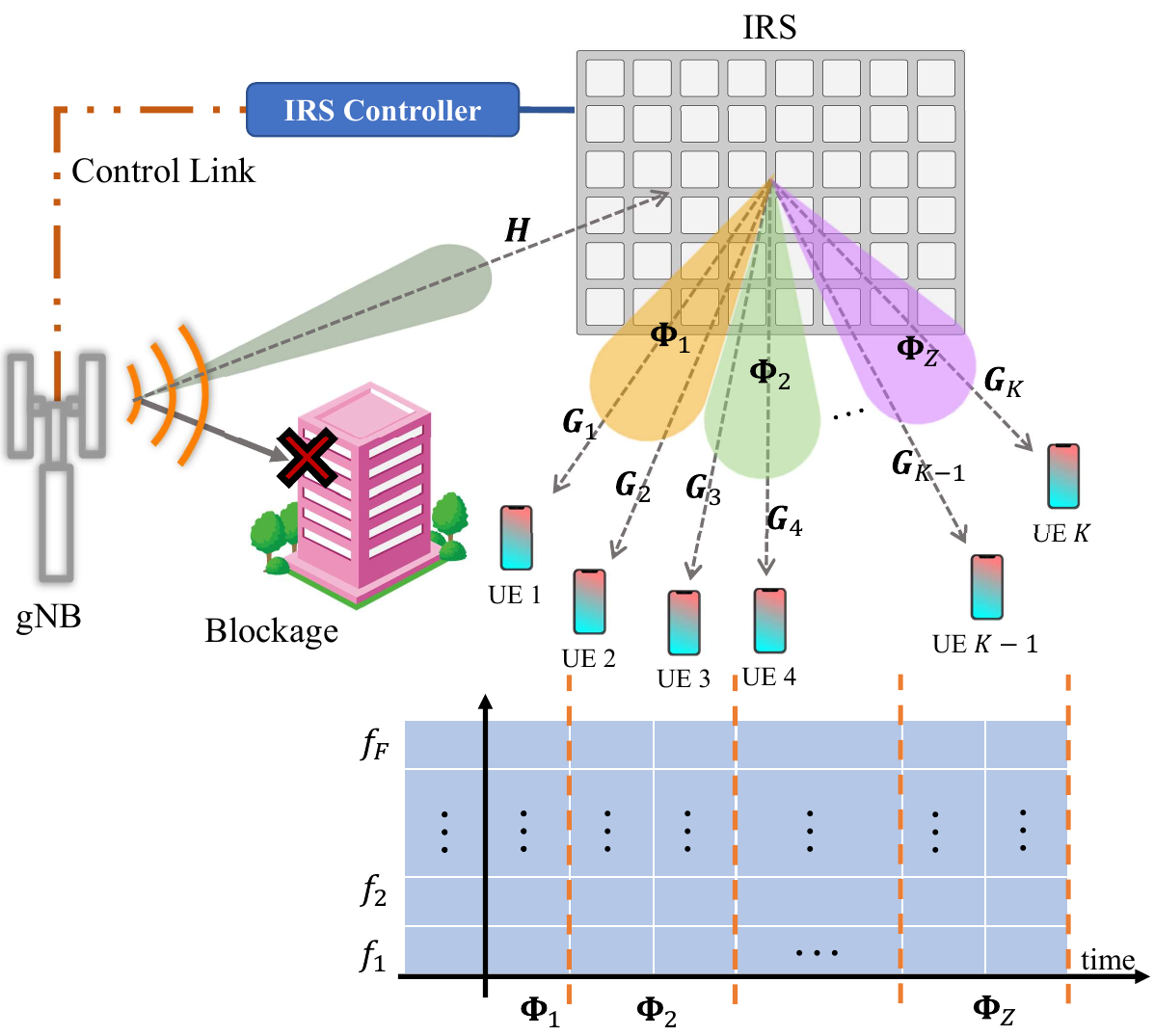}
    \caption{\ac{ofdma} scheduling for \ac{irs}-assisted multi-\ac{ue} communication.}
    \label{fig:system_model}
\end{figure}

We consider the downlink transmission of a cellular system shown in Fig.~\ref{fig:system_model}, where the transmission from the \ac{gnb} to  $K$ \acp{ue} is assisted by an \ac{irs}. 
The \ac{gnb} and each \acp{ue} are equipped with $N_{\rm g}$ and $N_{\rm U}$ antennas, respectively. We assume that the direct links between the \ac{gnb} and the \acp{ue} are unavailable due to a deep blockage, therefore the \ac{gnb} transmits signals to the \acp{ue} by only exploiting the \ac{irs} cascade channels.
The \ac{irs} configuration is managed by the \ac{gnb} though the \ac{irs} controller, by exploiting a dedicated link between the \ac{gnb} and the \ac{irs}.

The \ac{gnb} schedules the \acp{ue} in the time-frequency domain by allocating 
\acp{rb}
from a grid of $K/F$ \acp{rb}. Each \ac{ue} is assigned exactly one \ac{rb} in the considered resource grid, as we consider $K$ as an integer multiple of $F$, and all the \acs{ue} must be scheduled in each time frame. Let $f_i$ be the carrier frequency of the \ac{rb} identified by frequency index $i$ and an arbitrary time slot, and let $\mathcal F = \{f_i, i=1, \ldots, F\}$ be the set of all carrier frequencies.
We assume that \acp{ue} are either static or moving slowly, which is the most typical application scenario for \ac{irs}-aided networks. Therefore, we assume that once the perfect channel estimation of all \acp{ue} is acquired at the \ac{gnb} at the beginning of each frame, the channels remain constant for its duration. We assume that the gNB knows the cascade channel to the \acp{ue} for any \ac{irs} configuration. 

\vspace{5pt}\noindent\emph{\ac{irs} Model \& Beamforming Codebook.}
Each element of the \ac{irs} acts as an omnidirectional antenna element that reflects the impinging electromagnetic field, introducing a tunable phase shift on the baseband-equivalent signal. We denote with $\phi_n=e^{j\theta_n}$ the reflection coefficient of the $n$-th \ac{irs} element, where $\theta_n \in \left\{0, \frac{2\pi}{2^{b_{\rm I}}},\ldots,\frac{2\pi(2^{b_{\rm I}}-1)}{2^{b_{\rm I}}} \right\}$ is the induced phase shift, with $b_{\rm I}$-bits quantization.

To further reduce the overhead of the \ac{irs} configuration, we consider a configuration codebook $\mathcal{C}_{\bm{\Phi}}$ from which such matrix is chosen.
Such a discrete design is compliant with the currently standardized initial access framework~\cite{3gpp.38.214}. A large variety of codebooks for both near-field and far-field communication has been discussed in the literature, and the evaluation of their impact on system performance is out of the scope of this work. 
In this paper, we considered a simple design of \textit{cell-specific codebook}, which is derived from the channel measurements in the cell (the details are presented in Section~\ref{sec:codebook}).  

\vspace{5pt}\noindent\emph{Transmission model.}
With reference to carrier $i$, we denote with $\bm{H}(f_i) \in \mathbb{C}^{N_{\rm I} \times N_{\rm g}}$ the \ac{gnb}-\ac{irs} channel matrix and with $\bm{G}_k(f_i) \in \mathbb{C}^{N_{\rm U} \times N_{\rm I}}$ the channel matrix of the link between the \ac{irs} and \ac{ue} $k$.
We consider single-stream transmissions, where the \ac{gnb} uses the beamforming vector $\bm{w}_{\rm g} \in \mathbb{C}^{N_{\rm g}\times 1}$. Note that this assumption matches the \ac{irs}-aided \ac{mmwave} scenario, where the cascade channel rank is insufficient to perform multi-stream transmissions \cite{he2020cascaded, rains2023ris}.
Let $\bm{v}_k \in \mathbb{C}^{N_{\rm U}\times1}$ be the beamforming vector at \ac{ue} $k$, while $\bm{\Phi} = \diag(\phi_1,\ldots,\phi_{N_{\rm I}})$  is the {\em \ac{irs} configuration}. We assume that the \ac{gnb} beamforms the signal toward the \ac{gnb}-\ac{irs} \ac{los} angle and, once the \ac{irs} configuration is fixed. Then, the beamformer at the \ac{ue} matches its cascade channel, i.e., $\bm{v}_k$ is the singular vector corresponding to the largest singular value of $[\bm{G}_k(f_i) \bm{\Phi} \bm{H}(f_i)]^{\dagger}$. \ac{ue} $k$ attains the achievable rate
\begin{equation}\label{ar}
R_k(\bm{\Phi}, f_i) = \log_2\left(1+\frac{|\bm{v}_{k}^{\rm T}\bm{G}_k(f_i) \bm{\Phi} \bm{H}(f_i)  \bm{w}_{{\rm g}} |^2\sigma_{s}^2}{|\bm{v}_k^{\rm T}|^2\sigma_{n}^2}\right),
\end{equation}
where $\sigma_{s}^2$ and $\sigma_{n}^2$ are the transmit and noise signal power, respectively.

\section{Configuration and User Scheduling}\label{sec:scheduling}

In general, different \ac{irs} configurations should be adopted for each \ac{ue} to maximize its achievable rate \eqref{ar} based on its position in the cell and on the channel conditions. However, $\bm{\Phi}$ is not frequency-selective, i.e., its configuration is the same at each \ac{rb} in the same time slot.
Moreover, we here impose a constraint on the number of \ac{irs} reconfigurations per time frame, with the goal of either limiting the reconfiguration and reducing the overhead, or accounting for practical limitations that might arise in realistic deployments.
This limitation usually leads to an achievable rate degradation as sub-optimal \ac{irs} configurations could be adopted to serve some \acp{ue}. 

To mitigate this effect, we formulate a constrained discrete optimization problem, imposing a limit on the re-configurations within a time frame to a maximum number of $Z\leq K/F$. Within this time frame the \ac{gnb} serves the $K$ \acp{ue} by splitting them into $Z$ disjoint subsets (or clusters) $\mathcal{U}_1,\ldots,\mathcal{U}_{Z}$, each with cardinality $\alpha_z F$, with $\alpha_z \in \mathbb{N}$, for $z=1, \ldots, Z$, and assigning each \ac{ue} to one \ac{rb}.
While serving the \ac{ue} in subset $\mathcal{U}_{z}$ the \ac{irs} configuration is fixed to $\bm{\Phi}_z$.
Note that, if the codebook is small, several sets of \acp{ue} could share the optimal configuration; in such a case, the clusters are merged, and the number of reconfigurations is reduced.

Let $\mathcal{S}_i$, $i = 1,\ldots, F$, 
be the set of \acp{ue} assigned to carrier $i$. Also, define the assignment variables
\begin{equation}\label{assignment_variables}
    x_{k,z,i} = 
    \begin{cases}
        1 \quad \text{if } k \in \mathcal{U}_z \cap \mathcal{S}_i,\\
        0 \quad \text{otherwise}.
    \end{cases}  
\end{equation}
The joint resource allocation and configuration optimization problem can be stated as the following generalized assignment 
\begin{subequations}\label{optproblemMKP}
    \begin{equation}\label{fitfunction}
        \max_{\substack{\bm{\Phi}_z, \alpha_z \\ x_{k,z,i}}}\sum_{z = 1}^Z\sum_{k=1}^K\sum_{i=1}^F R_{k}(\bm{\Phi}_z, f_i) x_{k,z,i}
    \end{equation}
    \begin{alignat}{2}
     \text{s.t.}\quad & 
     \bm{\Phi}_z \in \mathcal{C}_{\bm{\Phi}},\label{constr_codebook}\\
      & x_{k,z,i} \in \{0,1\},\quad \forall k, z, i, \label{constr_01}\\
      &\sum_{z = 1}^Z\sum_{i=1}^Fx_{k,z,i} = 1 \quad \forall k, \label{constr_1}\\
      &\sum_{k=1}^Kx_{k,z,i} = \alpha_z \quad \forall i,z,\label{constalpha}
    \end{alignat}
\end{subequations}
where constraint \eqref{constr_codebook} imposes the \ac{irs} configurations to be chosen within codebook $\mathcal{C}_{\bm{\Phi}}$, and  \eqref{constr_01}-\eqref{constr_1} denote the assignment of each user to a unique \ac{rb}.
Instead, constraint \eqref{constalpha} imposes the cluster cardinalities as an integer multiple of $F$, reflecting the frequency non-selectivity of the \ac{irs} configuration.
Due to \eqref{constr_1}-\eqref{constalpha}, $\sum_{z=1}^Z \alpha_z = K/F$, and $\alpha_z$ is the number of time slots for which $\bm{\Phi}_z$ is kept. 

Note that \eqref{optproblemMKP} belongs to the class of generalized multi-knapsack problems, well-known as NP-complete. Its solution requires an exhaustive search over all the discrete parameters, therefore, a heuristic approach, which splits \eqref{optproblemMKP} into two sub-problems, is required. Moreover, we remark that the size of codebook $\mathcal{C}_{\bm{\Phi}}$ may be extremely large, up to the case $|\mathcal{C}_{\bm{\Phi}}|=2^{b_{\rm I}N_{\rm I}}$, i.e., all the combinations of phase shifts, thus exacerbating the problem complexity. 

\subsection{Optimization Problem Decomposition}
With the aim of simplifying problem \eqref{optproblemMKP}, we first propose to decompose the joint resource allocation and configuration assignment into two sub-problems named, respectively, the \textit{configurations assignment}, and the \textit{\ac{rb} assignment}. 

The configuration assignment sub-problem assigns one UE per cluster, leaving $K-Z$ \acp{ue} unassigned, and sets the \ac{irs} configuration for each cluster. The problem can be written as 
\begin{subequations}\label{subprob1}
    \begin{equation}
    \max_{\substack{\bm{\Phi}_z\\x_{k,z,i}}}\sum_{z=1}^Z\sum_{i=1}^F R_{k}(\bm{\Phi}_z, f_i) x_{k,z,i}
    \end{equation}
    \begin{alignat}{2}
     \text{s.t.}\quad & 
     \eqref{constr_codebook}, \eqref{constr_01}, \; 
     &\sum_{k = 1}^K\sum_{i=1}^Fx_{k,z,i} = 1 \quad\forall z.
      \end{alignat}
\end{subequations}

In the \ac{rb} assignment sub-problem, instead, the remaining \acp{ue} are assigned to the clusters defined with \eqref{subprob1} as 
    \begin{equation} \label{subprob2}
        \max_{\substack{\alpha_z\\ x_{k,z,i}}}\sum_{z = 1}^Z\sum_{k=1}^K\sum_{i=1}^F R_{k}(\bm{\Phi}_z, f_i) x_{k,z,i}, \;
     \text{s.t. }  \eqref{constr_01}, \eqref{constr_1}, \eqref{constalpha}.
\end{equation}
Note that both the \ac{rb} assignments and the cluster cardinality constraint \eqref{constalpha}, are assessed in this second step, as $\alpha_z$, $z = 1, \ldots, Z$  are optimization variables. 
Moreover, \eqref{subprob2} is still a multi-knapsack assignment problem with variable knapsack capacities, therefore belonging to the class of NP-complete problems.

\subsection{Greedy Maximum-Rate Scheduler (GMAX)}

Due to the complexity of \eqref{optproblemMKP}, we resort to a greedy approach to obtain a heuristic but close-to-optimal solution, by proposing the \ac{gmax} algorithm, summarized in Algorithm \ref{alg:gmax}. 

First, we observe that \eqref{subprob1}  can be solved by exhaustively computing $R_{k}(\bm{\Phi}, f_i)$ for all $i = 1, \ldots, F$, $\bm{\Phi} \in \mathcal{C}_{\bm{\Phi}}$, and $k = 1,\ldots, K$, and then selecting the $Z$ \acp{ue} (with their \ac{irs} configuration) providing the highest rate. Each of the selected \acp{ue} is assigned to \ac{rb} maximizing \eqref{ar}, respectively.


To handle the remaining \acp{ue} and solve \eqref{subprob2}, instead, we resort to a greedy 
approach.
Let $\mathcal{P} = \{\bm{\Phi}_1, \bm{\Phi}_2, \ldots, \bm{\Phi}_Z\}$ be the set of \ac{irs} configurations of each cluster, \ac{gmax} solves 
\begin{subequations}\label{secondITER}
    \begin{equation}
        (k, z, i)  = \argmax_{\substack{k, z, i}} R_{k}(\bm{\Phi}_z, f_i)
    \end{equation}
    \begin{alignat}{2}
     \text{s.t.}\quad & \eqref{constalpha},\,\bm{\Phi}_z\in \mathcal{P},\\
     &k \in \{k: x_{k,z,i}=0 \;\forall\,z,i\},\label{const_k}\\ 
    &(z,i) \in \{(z,i): x_{k,z,i}=0 \;\forall\,k\}.\label{const_zi}
    \end{alignat}
\end{subequations}
Since $Z\leq \frac{K}{F}$ in general, the \acp{ue} are firstly allocated considering only $F$ \acp{rb} per cluster, i.e., one-time slot per \ac{irs} configuration, by setting $\alpha_z = 1$ for all $z$.
Once the first $ZF$ \acp{ue} are allocated, the algorithm proceeds by solving problem \eqref{secondITER}, considering the allocation of new time slots in the resource grid (i.e. increasing $\alpha_z$ by one).

At the end of the procedure, each \ac{ue} is assigned to a specific \ac{rb}, satisfying all the constraints of problem \eqref{optproblemMKP}.
Note that, as per \eqref{assignment_variables}, sets $\mathcal{U}_1,\ldots,\mathcal{U}_{Z}$, and $\mathcal{S}_i$ are uniquely determined by variables $x_{k,z,i}$, for $z = 1\ldots, Z$, $i = 1,\ldots, F$, $k = 1,\ldots, K$.  

\subsection{Codebook Design And Control Overhead}\label{sec:codebook}
To obtain the cell-specific codebook of \ac{irs} configurations $\mathcal{C}_{\bm{\Phi}}$, a clustering-based approach is employed.
In particular, similarly to the distance-based clustering proposed in \cite{rech2023downlinkJ}, the points to be clustered are the \ac{irs} phase shifters (with $b_{\rm I}$-bits quantization) that maximize the achievable rate \eqref{ar}, at each $f\in \mathcal{F}$, of $M$ \acp{ue} deployed at random positions in the cell, with $M \gg K$.
Such configurations are grouped into $|\mathcal{C}_{\bm{\Phi}}|\ll M$ clusters, according to the well-known \ac{km} clustering \cite{kmeans}, and the codebook entries are the resulting cluster centroids.

Note that the codebook allows a substantial reduction of the \ac{irs} control link overhead. Indeed, for each \ac{irs} reconfiguration, the \ac{gnb} transmits $b_{\rm q} = \log_2|\mathcal{C}_{\bm{\Phi}}|$ bits, instead of the $b_{\rm I}N_{\rm I}$ bits needed to configure each phase shifter individually.
Moreover, by further limiting the number of reconfigurations per time frame to $Z$, the total number of control bits is reduced by a factor $\frac{ZF}{K}\leq 1$.

\begin{algorithm}[t]
\caption{Greedy Maximum-Rate Scheduler}\label{alg:gmax}
\begin{algorithmic}[1]
\STATE \textbf{Input:} $R_{k}(\bm{\Phi}, f_i) \text{ for all } k, i, \bm{\Phi}\in\mathcal{C}_{\bm{\Phi}}$
\STATE \textbf{Output:} $\mathcal{P}, x_{k, z, i}, \text{ for all } k, z, i$
\STATE $\alpha_z \gets 1, \quad \text{ for all } z$
\STATE $x_{k, z, i} \gets 0, \quad \text{ for all } k, z, i$
\STATE $(x_{k, z, i}, \bm{\Phi}_z)\gets$ solve \eqref{subprob1} exhaustively
\WHILE{$\sum_{z = 1}^Z\sum_{k=1}^K\sum_{i=1}^Fx_{k,z,i} < K$}
    \WHILE{$\sum_{z = 1}^Z\sum_{k=1}^K\sum_{i=1}^Fx_{k,z,i} \leq F \sum_{z=1}^Z \alpha_z$}
        \STATE $x_{k, z, i} \gets 1\quad \text{ for } k, i \text{ solving \eqref{secondITER}}$
    \ENDWHILE
    \IF{$\sum_{z = 1}^Z\sum_{k=1}^K\sum_{i=1}^Fx_{k,z,i} < K$}
        \STATE $(k,z,i)\gets$ solve \eqref{secondITER} neglecting constraint \eqref{constalpha}
        \STATE $\alpha_z\gets\alpha_z+1$
        \STATE $x_{k, z, i} \gets 1$    
    \ENDIF    
\ENDWHILE
\end{algorithmic}
\end{algorithm}

\subsection{Computational Complexity}
The computational complexity of \ac{gmax} is mainly due to the maximum rate computation for the initial choice of the $Z$ \ac{irs} configurations to solve \eqref{subprob1}. 
Specifically, the cascade channel matrix product $\bm{G}_k \bm{\Phi}_k \bm{H}$ dominates computations with a complexity $O\big(N_{\rm g}N_{\rm I}^2 + N_{\rm g}N_{\rm I}N_{\rm U}\big)$, and the procedure must be done for all \acp{ue}, carrier frequencies, and \ac{irs} configurations in the codebook. 
Similarly, the second loop computes $R_k$ in the same fashion, but the search is restricted to set $\mathcal{P}$, and typically $|\mathcal{P}| = Z\ll2^{b_{\rm q}}$.
As a result, the overall complexity of \ac{gmax} is $O\big(ZF(2^{b_{\rm q}}+ 1)(N_{\rm g}N_{\rm I}^2 + N_{\rm g}N_{\rm I}N_{\rm U})\big)$.
Note that, in the first step the complexity grows exponentially with the codebook overhead, penalizing codebooks of large resolution. This suggests the adoption of a cell-specific codebook to maximize the rate with low overhead.   
However, a further complexity reduction can be achieved by observing that, in the first loop, only the optimal \ac{irs} configuration of each \ac{ue}, i.e., the one maximizing its transmission rate, is needed.
A possible approach is to derive the optimal \ac{irs} configuration $\bm{\Phi}_k^{'*}(f_i)$ for all $k, i$ in the continuous phase domain. $\bm{\Phi}_k^{'*}(f_i)$ is then mapped to the closest (in the sense of circular distance~\cite{rech2023downlinkJ}) codeword in the codebook $\bm{\Phi}_k^*\in\mathcal{C}_{\bm \Phi}$.
While the time complexity of deriving $\bm{\Phi}_k^{'*}(f_i)$ for each $k, i$ is $O\big(N_{\rm g}N_{\rm I}^2 + N_{\rm g}N_{\rm I}N_{\rm U}\big)$~\cite{rech2023downlinkJ}, its approximation requires $O(2^{b_{\rm q}} N_{\rm I})$ operations. 
With this approximation, the total complexity can be reduced to $O\big(K(2^{b_q} + (N_{\rm g}N_{\rm I}^2 + N_{\rm g}N_{\rm I}N_{\rm U}N_{\rm I})) + ZF(N_{\rm g}N_{\rm I}^2 + N_{\rm g}N_{\rm I}N_{\rm U})\big)$.

\section{Numerical Results}\label{sec:numerical_results}
We consider the \ac{umi} \ac{3gpp} scenario~\cite{3gpp.38.901}, with all devices lying in the 2-D plane with the \ac{gnb} placed at the center. 
According to the \ac{3gpp} specifications, the coverage area of the \ac{gnb} is characterized by an average radius of $167$~m and is assumed to lie in the positive $x$-axis region.
We consider $K=90$ \acp{ue} are randomly deployed according to a uniform distribution within the cell area, to be served in downlink by the \ac{gnb}, assisted by an \ac{irs} at coordinates  $(75, 100)$~m. 
The \ac{gnb} and the \acp{ue} are equipped with \ac{ula} with $N_{\rm g}=32$ and $N_{\rm U} = 4$ antennas, while for the \ac{irs}, if not otherwise specified, we adopt a $20$H$\times40$V reflective panel ($N_{\rm I}= 800$), $b_{\rm I}=1$ phase shift quantization bits, and $b_{\rm q}=14$ bits for the codebook overhead.

\vspace{5pt}\noindent\emph{Channel.}
The system operates at a carrier frequency of $f_{\rm c}=28$~GHz, the \ac{gnb} transmission power is $33$~dBm, and the noise power spectral density at the receivers is $-174$~dBm/Hz. \acp{rb} are equally spaced in the band $(f_{\rm c} - 10~{\rm MHz}, f_{\rm c} + 10~{\rm MHz})$.
We consider the \ac{3gpp} TR~38.901 spatial channel model~\cite{3gpp.38.901}, wherein channel matrices are computed based on the superposition of different clusters, each consisting of multiple rays that arrive (depart) to (from) the antenna arrays with specific angles and powers. The link between \ac{gnb} and \acp{ue}  experiences deep blockage, while we consider a \ac{los} link between the \ac{gnb} and the \ac{irs}, and the channels between \ac{irs} and \acp{ue} may exhibit a \ac{los} component depending on their distance, according to~\cite{3gpp.38.901}.


\begin{table} 
\centering
\caption{Average Sum Rate for Different IRS Sizes ${\rm[bit/s/Hz]}$}
\label{tab:tab_codebook}
\begin{tabular}{lcccc}
\toprule
     &  $b_{\rm q}=12$ & $b_{\rm q}=14$ & $b_{\rm q}=16$ & $b_{\rm q}=N_{\rm I}$ \\\midrule
$10$H$\times20$V & 12.89 & 19.06 & 20.92 & 21.36 \\ 
$20$H$\times40$V & 44.65 & 69.97 & 78.33 & 92.16 \\ 
$30$H$\times60$V & 86.90 & 108.33 & 131.65 & 162.19 \\
 \bottomrule
\end{tabular}
\end{table}

The system performance is evaluated in terms of \textit{average sum rate}, defined as 
\begin{equation}
    \bar{R} = \mathbb{E}\left[\sum_{z = 1}^Z\sum_{k=1}^K\sum_{i=1}^F R_{k}(\bm{\Phi}_z, f_i)x_{k,z,i}\right],
\end{equation}
where we average over multiple channel realizations and randomly generated \acp{ue} positions.

\subsection{Compared Solutions}
In the following, we compare \ac{gmax} with different resource allocation policies, under different codebook sizes.

\noindent\emph{\Ac{da}.} As a baseline, in this scheduling each \ac{ue} is deterministically assigned to an \ac{rb} in cluster $z$ and, upon the assignment, the \ac{irs} configuration $\bm{\Phi}_z\in \mathcal{C}_{\bm{\Phi}}$ maximizes the cluster sum rate.

\noindent\emph{\Ac{uoscbc}.} This is an extension to \ac{ofdma} scheduling of the \ac{oscbc} proposed in \cite{rech2023downlinkJ} for \ac{tdma}. The unique assignment to a particular \ac{rb}, i.e., constraint \eqref{constr_1} is violated, as there is no limitation imposed in the number of \acp{ue} associated with each \ac{rb}.


\noindent\emph{\Ac{ga}.} This is a \ac{ga} \cite{holland1992genetic} with fitness function \eqref{fitfunction}, whose initial population includes the \ac{gmax} solution. In such \ac{ga} approach the population generation, crossover, and mutation functions are customized such that all the constraints \eqref{constr_codebook}-\eqref{constalpha} are always satisfied. This provides the (almost) optimal solution of problem \eqref{optproblemMKP}.

\subsection{Performance Results}
Firstly, Table~\ref{tab:tab_codebook} investigates the relationship between codebook size and system sum rate in the ideal case with $F = 1$ and each \ac{ue} scheduled with its optimal \ac{irs} configuration $\bm{\Phi}^*_k$. 
The results reveal the need for a large codebook to approximate the continuous case (i.e., $b_{\rm q} = N_{\rm I}$ for $b_{\rm I}=1$).
Also, larger \ac{irs} panels are more sensitive to the codebook size, 
as a result of the larger number of degrees of freedom provided by the independent control of each phase shifter.
For example, a $10$H$\times20$V-elements \ac{irs} achieves around $60$\% of the sum-rate achievable with the continuous codebook with only $b_{\rm q}=12$, and $98$\%  for $b_{\rm q}=16$. Instead, a $30$H$\times60$V-elements \ac{irs} requires $b_{\rm q}=16$ to reach $81$\% of the sum rate achievable in the continuous case.

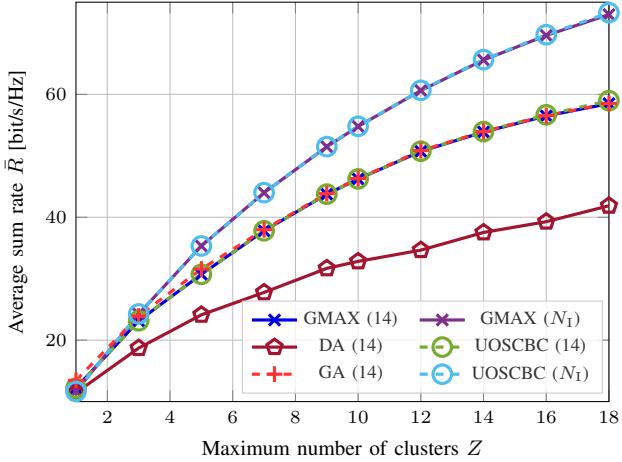
\begin{figure}
    \centering
       \setlength\fwidth{0.8\columnwidth}
    \setlength\fheight{0.6\columnwidth}
    \definecolor{colorGMAX}{rgb}{0,0,0.9}
\definecolor{colorCGMAX}{rgb}{0.49400,0.18400,0.55600}
\definecolor{colorGA}{rgb}{1,0.2,0.2}
\definecolor{colorUOSCBC}{rgb}{0.46600,0.67400,0.18800}%
\definecolor{colorCUOSCBC}{rgb}{0.30100,0.74500,0.93300}%
\definecolor{colorOSCBCC}{rgb}{0.92900,0.69400,0.12500}%
\definecolor{colorDA}{rgb}{0.63500,0.07800,0.18400}%

\pgfplotsset{every tick label/.append style={font=\scriptsize}}
\begin{tikzpicture}

\begin{axis}[%
    width=\fwidth,
    height=\fheight,
    at={(0\fwidth,0\fheight)},
    scale only axis,
    xlabel style={font=\footnotesize},
    ylabel style={font=\footnotesize},
    xmin=1,
    xmax=18,
    ymin=10,
    ymax=75,
    xlabel={Maximum number of clusters $Z$},
    ylabel={Average sum rate $\bar{R}$ [bit/s/Hz]},
    axis background/.style={fill=white},
    xmajorgrids,
    ymajorgrids,
        legend image post style={mark indices={}},
    legend style={
        /tikz/every even column/.append style={column sep=0.2cm},
        at={(0.65, 0.02)}, 
        anchor=south, 
        draw=white!80!black, 
        font=\scriptsize,
        fill opacity=0.8
        },
    legend columns=2,
]

\addplot [color=colorGMAX, very thick, mark size=3pt, mark=x, mark options={solid, colorGMAX}]
  table[row sep=crcr]{%
1	12.2032945087665\\
3	23.1497334656637\\
5	30.7123299390868\\
7	37.7908455856924\\
9	43.7513106116052\\
10	46.2247950332475\\
12	50.7164089843852\\
14	53.8753158946102\\
16	56.4913520854847\\
18	58.4907903109734\\
};
\addlegendentry{GMAX (14)}

\addplot [color=colorCGMAX, very thick, mark size=3pt, mark=x, mark options={solid, rotate=180, colorCGMAX}]
  table[row sep=crcr]{%
1	11.6432230412029\\
3	24.2612098442878\\
5	35.3257652278178\\
7	43.992281711625\\
9	51.4733420349904\\
10	54.7820443615611\\
12	60.6210138456496\\
14	65.5920222263067\\
16	69.5878908788619\\
18	72.9877742155636\\
};
\addlegendentry{GMAX ($N_{\rm I}$)}

\addplot [color=colorDA, very thick, mark size=3pt, mark=pentagon, mark options={solid, colorDA}]
  table[row sep=crcr]{%
1	11.4334628353742\\
3	18.7079155713144\\
5	24.1212361866034\\
7	27.7667088436988\\
9	31.681173520335\\
10	32.8133991323814\\
12	34.6484201452414\\
14	37.5255830345727\\
16	39.2787094095557\\
18	41.866519661664\\
};
\addlegendentry{DA (14)}

\addplot [color=colorUOSCBC, dashed, very thick, mark size=3.5pt, mark=o, mark options={solid, colorUOSCBC}]
  table[row sep=crcr]{%
1	12.2032710104607\\
3	23.150518041397\\
5	30.7148225592508\\
7	37.7952316746478\\
9	43.7595560767972\\
10	46.2368035798152\\
12	50.7524096642566\\
14	53.9461658241532\\
16	56.6629664958839\\
18	58.9723951980875\\
};
\addlegendentry{UOSCBC (14)}

\addplot [color=colorGA, dashed, very thick, mark size=3pt, mark=+, mark options={solid, colorGA}]
  table[row sep=crcr]{%
1	13.442647288252\\
3	23.9220727901397\\
5	31.5707784752168\\
7	37.9763453763005\\
9	43.8454009817165\\
10	46.2964726245532\\
12	50.8544585989403\\
14	53.9527362689975\\
16	56.5316366059045\\
18	58.5223514837183\\
};
\addlegendentry{GA (14)}

\addplot [color=colorCUOSCBC, dashed, very thick, mark size=3.5pt, mark=o, mark options={solid, colorCUOSCBC}]
  table[row sep=crcr]{%
1	11.6441046038109\\
3	24.2645196705791\\
5	35.3326414964168\\
7	44.003886711114\\
9	51.4903735643194\\
10	54.8028360984966\\
12	60.6560277957905\\
14	65.6530919749727\\
16	69.7037275502913\\
18	73.2914939845402\\
};
\addlegendentry{UOSCBC ($N_{\rm I}$)}

\end{axis}

\end{tikzpicture}
    \caption{Average sum rate versus the number of clusters, for $K = 90$ \acp{ue}, $F = 5$ carriers. Between brackets is the number $b_{\rm q}$ of bits in the codebook.}
    \label{fig:sumcap_comp}
\end{figure}

\begin{figure}
    \centering
       \setlength\fwidth{0.8\columnwidth}
    \setlength\fheight{0.6\columnwidth}
    \input{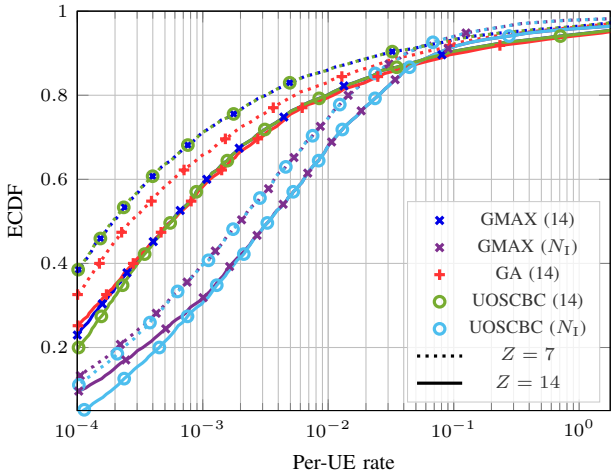}
    \caption{ECDF of the per-user rate, for $K = 90$ \acp{ue}, $F = 5$ carrier frequencies, and $Z\in\{7,14\}$. Between brackets is the number $b_{\rm q}$ of bits in the codebook.}
    \label{fig:cdf}
\end{figure}

Fig.~\ref{fig:sumcap_comp} depicts the average sum rate as a function of the number of clusters $Z$, comparing the different clustering strategies. Since each \ac{ue} must be scheduled once in the resource grid, $Z$ is bounded by $K/F$. The results show a huge performance gap between the proposal and the \ac{da} baseline and highlight the huge performance degradation due to the codebook resolution compared to the slight impact of the frequency assignment constraints \eqref{constr_1}-\eqref{constalpha}. In particular, \ac{gmax} and \ac{uoscbc} with high-resolution codebook ($b_{\rm q} = N_{\rm I}$) show a substantial sum rate gap with their respective low-resolution codebook case ($b_{\rm q} = 14$), while the negligence of constraints \eqref{constr_1}-\eqref{constalpha} with \ac{uoscbc} does not provide any substantial benefit on the performance. 
Note that the proposed \ac{gmax} schedulers perform very close to the \ac{ga}, which is shown to deviate very slightly from the greedy solution. Even though the \ac{ga} approach is not always optimal, such a negligible gap is representative of the validity of \ac{gmax} in this context. 
To emphasize the performance gap between the compared schemes,  Fig.~\ref{fig:cdf} shows the \ac{ecdf} of the per-\ac{ue} rate for fixed numbers of clusters $Z \in \{7,14\}$. Note that, while the performance hierarchy remains invariant for almost all compared schemes, the adoption of the sum rate as the fitness function of \ac{ga} may result in a different rate distribution than \ac{gmax}, promoting the \acp{ue} experiencing the best channel conditions while penalizing the others.

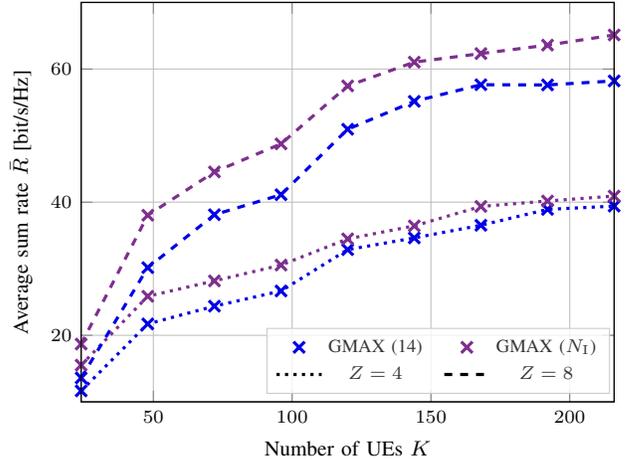
\begin{figure}
    \centering
       \setlength\fwidth{0.8\columnwidth}
    \setlength\fheight{0.6\columnwidth}
    \definecolor{colorGMAX}{rgb}{0,0,0.9}
\definecolor{colorCGMAX}{rgb}{0.49400,0.18400,0.55600}

\pgfplotsset{every tick label/.append style={font=\scriptsize}}
\begin{tikzpicture}

\begin{axis}[%
    width=\fwidth,
    height=\fheight,
    at={(0\fwidth,0\fheight)},
    scale only axis,
    xlabel style={font=\footnotesize},
    ylabel style={font=\footnotesize},
    xmin=24,
    xmax=216,
    ymin=10,
    ymax=70,
    xlabel={Number of \acp{ue} $K$},
    ylabel={Average sum rate $\bar{R}$ [bit/s/Hz]},
    axis background/.style={fill=white},
    xmajorgrids,
    ymajorgrids,
        legend image post style={mark indices={}},
    legend style={
        /tikz/every even column/.append style={column sep=0.2cm},
        at={(0.55, 0.02)}, 
        anchor=south, 
        draw=white!80!black, 
        font=\footnotesize
        },
    legend columns=2,
]

\addplot [color=colorGMAX, dotted, very thick, mark size=3pt, mark=x, mark options={solid, colorGMAX}]
  table[row sep=crcr]{%
24	11.6512553898759\\
48	21.7116386408853\\
72	24.3788489566165\\
96	26.6575823144249\\
120	32.8662311083821\\
144	34.6270168029455\\
168	36.5278687561321\\
192	38.9145299715394\\
216	39.4073059128989\\
};
\addlegendentry{$Z=4$ GMAX}

\addplot [color=colorCGMAX, dotted, very thick, mark size=3pt, mark=x, mark options={solid, rotate=180, colorCGMAX}]
  table[row sep=crcr]{%
24	15.5214112331619\\
48	25.8644750928237\\
72	28.1743763952726\\
96	30.5506364571854\\
120	34.4855607111012\\
144	36.4357899253885\\
168	39.3576729305581\\
192	40.1742692315994\\
216	40.9142742720383\\
};
\addlegendentry{$Z=4$ C-GMAX}

\addplot [color=colorGMAX, very thick, dashed, mark size=3pt, mark=x, mark options={solid, colorGMAX}]
  table[row sep=crcr]{%
24	13.6015202253067\\
48	30.1665440688855\\
72	38.1160380331104\\
96	41.119019636103\\
120	50.9531298246056\\
144	55.1453008914095\\
168	57.6413282949928\\
192	57.6091277377713\\
216	58.2142478609558\\
};
\addlegendentry{$Z=8$ GMAX}

\addplot [color=colorCGMAX, very thick, dashed, mark size=3pt, mark=x, mark options={solid, rotate=180, colorCGMAX}]
  table[row sep=crcr]{%
24	18.6883962670917\\
48	38.0342797525047\\
72	44.544174419547\\
96	48.7574216121042\\
120	57.4610400068288\\
144	61.0356427277516\\
168	62.3202683035146\\
192	63.5921308393077\\
216	65.1138877080558\\
};
\addlegendentry{$Z=8$ C-GMAX}

\legend{}
\end{axis}

\begin{axis}[%
    width=\fwidth,
    height=\fheight,
    at={(0\fwidth,0\fheight)},
    scale only axis,
    xmin=1,
    xmax=100,
    xtick={},
    ytick={},
    xticklabels={{}, {}, {},{}},
    yticklabels={},
    xtick style = {draw=none},
    ytick style = {draw=none},
    ymin=0.0001,
    ymax= 1,
    legend style={
            /tikz/every even column/.append style={column sep=0.2cm},
            at={(0.67, 0.02)}, 
            anchor=south, 
            draw=white!80!black, 
            font=\scriptsize,
            fill opacity=0.8
            },
        legend columns=2,
    ]

\addplot [color=colorGMAX, very thick, only marks, mark size=3pt, mark=x, mark options={solid, colorGMAX}]
  table[row sep=crcr]{%
1	-5 \\
};
\addlegendentry{GMAX (14)}

\addplot [color=colorCGMAX, very thick, only marks, mark size=3pt, mark=x, mark options={solid, rotate=180, colorCGMAX}]
  table[row sep=crcr]{%
1	-5 \\
};
\addlegendentry{GMAX ($N_{\rm I}$)}

\addplot [color=black, dotted, very thick]
  table[row sep=crcr]{%
1	-5\\
};
\addlegendentry{$Z=4$}

\addplot [color=black, dashed, very thick]
  table[row sep=crcr]{%
1	-5\\
};
\addlegendentry{$Z=8$}

\end{axis}

\end{tikzpicture}
    \caption{Average sum rate versus the number of \acp{ue}, for $F = 3$ carrier frequencies, and $Z\in\{4,8\}$. Between brackets is the number $b_{\rm q}$ of bits in the codebook.}
    \label{fig:sumcap_vsK}
\end{figure}

Fig.~\ref{fig:sumcap_vsK} shows the average sum rate as a function of $K$, for $F=3$ carrier frequencies. While the sum rate increases with $K$, for low numbers of clusters ($Z=4$) the performance gap between \ac{gmax} with the low-resolution codebook and \ac{gmax} with $b_{\rm q} = N_{\rm I}$ becomes negligible for a large number of \acp{ue}, as the configurations associated to each cluster are sub-optimal in maximizing the sum rate in both cases.

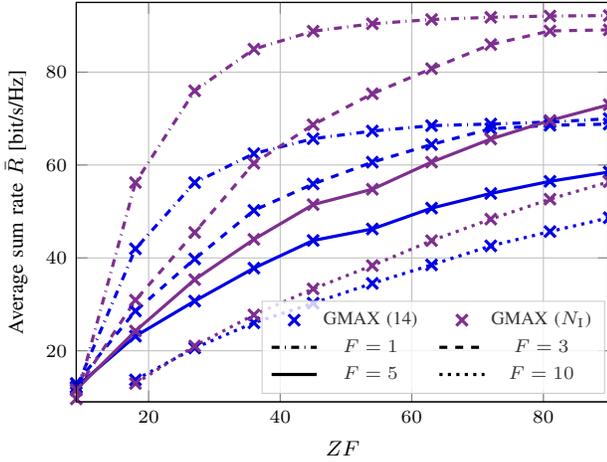
\begin{figure}
    \centering
       \setlength\fwidth{0.8\columnwidth}
    \setlength\fheight{0.6\columnwidth}
    \definecolor{colorGMAX}{rgb}{0,0,0.9}
\definecolor{colorCGMAX}{rgb}{0.49400,0.18400,0.55600}

\pgfplotsset{every tick label/.append style={font=\scriptsize}}
\begin{tikzpicture}

\begin{axis}[%
    width=\fwidth,
    height=\fheight,
    at={(0\fwidth,0\fheight)},
    scale only axis,
    xlabel style={font=\footnotesize},
    ylabel style={font=\footnotesize},
    xmin=9,
    xmax=90,
    ymin=9,
    ymax=95,
    xlabel={$ZF$},
    ylabel={Average sum rate $\bar{R}$ [bit/s/Hz]},
    axis background/.style={fill=white},
    xmajorgrids,
    ymajorgrids,
        legend image post style={mark indices={}},
    legend style={
        /tikz/every even column/.append style={column sep=0.2cm},
        at={(0.55, 1.02)}, 
        anchor=south, 
        draw=white!80!black, 
        font=\footnotesize
        },
    legend columns=2,
]

\addplot [color=colorGMAX, dashdotted, very thick, mark size=3pt, mark=x, mark options={solid, colorGMAX}]
  table[row sep=crcr]{%
9	11.9867561719265\\
18	41.9588360232641\\
27	56.212278054767\\
36	62.454108544354\\
45	65.6675300549675\\
54	67.3118483083442\\
63	68.4764907401639\\
72	68.8254737149043\\
81	69.2525923917826\\
90	69.9711360481164\\
};
\addlegendentry{GMAX $F = 1$}

\addplot [color=colorCGMAX, dashdotted, very thick, mark size=3pt, mark=x, mark options={solid, rotate=180, colorCGMAX}]
  table[row sep=crcr]{%
9	9.64071694962127\\
18	56.2149903325459\\
27	75.9780502838299\\
36	84.9553394393348\\
45	88.8081462720178\\
54	90.412530980734\\
63	91.327414773895\\
72	91.8018519395597\\
81	92.0773595663118\\
90	92.1694584647754\\
};
\addlegendentry{C-GMAX $F = 1$}

\addplot [color=colorGMAX, dashed, very thick, mark size=3pt, mark=x, mark options={solid, colorGMAX}]
  table[row sep=crcr]{%
9	12.9532918312959\\
18	28.5707557000478\\
27	39.7682524158478\\
36	50.2301910267205\\
45	55.9184290055693\\
54	60.6086985083816\\
63	64.407267097842\\
72	67.8058828439543\\
81	68.5498675223362\\
90	68.8088419959204\\
};
\addlegendentry{GMAX $F = 3$}

\addplot [color=colorCGMAX, dashed, very thick, mark size=3pt, mark=x, mark options={solid, rotate=180, colorCGMAX}]
  table[row sep=crcr]{%
9	11.5372142450567\\
18	30.9034194066826\\
27	45.4768195527982\\
36	60.3734139400263\\
45	68.6707286348577\\
54	75.3476484048276\\
63	80.7528640518873\\
72	85.935246618591\\
81	88.86517350854\\
90	89.0989305942415\\
};
\addlegendentry{C-GMAX $F = 3$}

\addplot [color=colorGMAX, very thick, mark size=3pt, mark=x, mark options={solid, colorGMAX}]
  table[row sep=crcr]{%
9	12.2032945087665\\
18	23.1497334656637\\
27	30.7123299390868\\
36	37.7908455856924\\
45	43.7513106116052\\
54	46.2247950332475\\
63	50.7164089843852\\
72	53.8753158946102\\
81	56.4913520854847\\
90	58.4907903109734\\
};
\addlegendentry{GMAX $F = 5$}

\addplot [color=colorCGMAX, very thick, mark size=3pt, mark=x, mark options={solid, rotate=180, colorCGMAX}]
  table[row sep=crcr]{%
9	11.6432230412029\\
18	24.2612098442878\\
27	35.3257652278178\\
36	43.992281711625\\
45	51.4733420349904\\
54	54.7820443615611\\
63	60.6210138456496\\
72	65.5920222263067\\
81	69.5878908788619\\
90	72.9877742155636\\
};
\addlegendentry{C-GMAX $F = 5$}

\addplot [color=colorGMAX, dotted, very thick, mark size=3pt, mark=x, mark options={solid, colorGMAX}]
  table[row sep=crcr]{%
18	13.7286322963348\\
27	20.6038976376384\\
36	26.0244442084365\\
45	30.2493355968588\\
54	34.5132026599637\\
63	38.4940985129512\\
72	42.588286584215\\
81	45.6666306294927\\
90	48.653382990205\\
};
\addlegendentry{GMAX $F = 10$}

\addplot [color=colorCGMAX, dotted, very thick, mark size=3pt, mark=x, mark options={solid, rotate=180, colorCGMAX}]
  table[row sep=crcr]{%
18	12.9536031733431\\
27	21.0348721227769\\
36	27.7281746952127\\
45	33.3380595404643\\
54	38.3449100485716\\
63	43.7152907732134\\
72	48.3446818174851\\
81	52.6382367765546\\
90	56.4044758632177\\
};
\addlegendentry{C-GMAX $F = 10$}

\legend{}
\end{axis}

\begin{axis}[%
    width=\fwidth,
    height=\fheight,
    at={(0\fwidth,0\fheight)},
    scale only axis,
    xmin=1,
    xmax=100,
    xtick={},
    ytick={},
    xticklabels={{}, {}, {},{}},
    yticklabels={},
    xtick style = {draw=none},
    ytick style = {draw=none},
    ymin=0.0001,
    ymax= 1,
    legend style={
            /tikz/every even column/.append style={column sep=0.2cm},
            at={(0.67, 0.02)}, 
            anchor=south, 
            draw=white!80!black, 
            font=\scriptsize,
            fill opacity=0.8
            },
        legend columns=2,
    ]

\addplot [color=colorGMAX, very thick, only marks, mark size=3pt, mark=x, mark options={solid, colorGMAX}]
  table[row sep=crcr]{%
1	-5 \\
};
\addlegendentry{GMAX (14)}

\addplot [color=colorCGMAX, very thick, only marks, mark size=3pt, mark=x, mark options={solid, rotate=180, colorCGMAX}]
  table[row sep=crcr]{%
1	-5 \\
};
\addlegendentry{GMAX ($N_{\rm I}$)}

\addplot [color=black, dashdotted, very thick]
  table[row sep=crcr]{%
1	-5\\
};
\addlegendentry{$F=1$}

\addplot [color=black, dashed, very thick]
  table[row sep=crcr]{%
1	-5\\
};
\addlegendentry{$F=3$}

\addplot [color=black, solid, very thick]
  table[row sep=crcr]{%
1	-5\\
};
\addlegendentry{$F=5$}

\addplot [color=black, dotted, very thick]
  table[row sep=crcr]{%
1	-5\\
};
\addlegendentry{$F=10$}
\end{axis}

\end{tikzpicture}
    \caption{Average sum rate versus $ZF$, for $K = 90$ \acp{ue}, $F \in \{1, 3, 5, 10\}$ carrier frequencies. Between brackets is the number $b_{\rm q}$ of bits in the codebook.}
    \label{fig:sumcap_vsFZ}
\end{figure}

Finally, to analyze the impact of the number of carrier frequencies, Fig.~\ref{fig:sumcap_vsFZ} shows the sum rate as a function of and $ZF$. Since $1\leq Z \leq KF$, the best performance is achievable for fewer carriers, allowing for more frequent reconfigurations. Moreover, it is shown that for large $Z$ the cases $F=1$ and $F=3$ exhibit very similar performance. This peculiar behavior is a direct consequence of the considered \ac{umi} cell, as $\sim33$\% of the \acp{ue} on average exhibit a \ac{los} channel component.
The channel gain experienced by such \acp{ue} is significantly larger than the gains of the \ac{ue} in \ac{nlos}. For $Z=K/3$, such users are allocated in different clusters and their optimal configurations are therefore chosen to serve their respective clusters. Thus, $Z=K/3$ is enough to obtain a high sum-rate performance.     

\section{Conclusions}\label{sec:conclusions}
We have discussed the \ac{ofdma} downlink scheduling in an \ac{irs}-assisted multi-user \ac{mimo} system, considering a limited number of \ac{irs} reconfigurations per time frame and a discrete codebook of possible configurations. We have tackled the sum rate maximization as an NP-complete generalized multi-knapsack problem, proposing a heuristic solution for the joint \ac{irs} configuration and resource allocation and showing its effectiveness in guaranteeing close-to-maximum sum rate compared to a \ac{ga}-based approach.

\balance
\bibliographystyle{IEEEtran}
\bibliography{IEEEabrv,biblio_new}

\end{document}